\documentclass[aps,prl,reprint,showpacs]{revtex4-1}
\usepackage{amsmath,amsthm,amsfonts,amssymb}

\usepackage{tikz}
\usetikzlibrary{arrows,calc,shapes.arrows}

\newtheorem{thm}{Theorem}

\newtheorem{lemma}{Lemma}
\newcommand {\tsf} [1]{\textsf{#1}}

\DeclareMathOperator{\tr}{tr}
\DeclareMathOperator{\ran}{ran}
\DeclareMathOperator{\supp}{supp}

\newcommand {\cl}{\mathcal}
\newcommand\mbf[1]{\mathbf{#1}}
\newcommand\ovl[1]{\overline{#1}}

\newcommand {\ket}[1] {|{#1}\rangle}
\newcommand {\kets} [1] {|#1\rangle_S}
\newcommand {\keti} [1] {|#1\rangle_I}
\newcommand {\ketsi} [1] {|#1\rangle_{IS}}
\newcommand {\brasi} [1] {\langle#1|}
\newcommand {\kete} [1] {|#1\rangle_E}
\newcommand {\bra}[1] {\langle{#1}|}

\newcommand{\braket}[2]{\langle{#1}|{#2}\rangle}

\begin{document}
\title{Optimal Quantum States for Image Sensing in Loss}
\author{Ranjith Nair and Brent J.~Yen}
\affiliation{Research Laboratory of Electronics, Massachusetts Institute of Technology, Cambridge, Massachusetts 02139, USA}
\begin{abstract} We consider a general image sensing framework that includes many quantum sensing problems by appropriate choice of image set, prior probabilities, and cost function. For any such problem, in the presence of loss and a signal energy constraint, we show that a pure input state of light with the signal modes in a mixture of number states minimizes the cost among all ancilla-assisted parallel strategies. Lossy binary phase discrimination with a peak photon number constraint and general lossless image sensing are considered as examples. \end{abstract}
\pacs{42.50.Ex, 42.50.Dv, 06.20.-f, 03.67.Hk}
\maketitle

The use of nonclassical and entangled states of light, i.e., states other than the easily generated coherent states and their classical mixtures \cite{classical}, for applications such as sub-shot-noise  imaging \cite{Brida10} and imaging with sub-Rayleigh resolution \cite{subrayleigh1,*subrayleigh2,*subrayleigh3} has received much attention. In the areas of sensing and metrology \cite{GLM11}, there have been recent theoretical studies of quantum-enhanced target detection \cite{Tan08}, reading of a digital memory \cite{Pirandola11,Nair11}, and of optical phase estimation \cite{DD1,*DD2} with nonclassical states. Given the interest in applications of quantum states of light for sensing, it is important to theoretically establish what state(s) accomplish a sensing task using minimum energy, assuming the most general measurements and post-processing. This would place a limit on the enhancements obtainable from nonclassical states using experimentally realizable measurements. The ubiquitous linear loss is known to be a bottleneck for harnessing quantum advantage in many communication and metrology applications \cite{Yuen04,DD1}. Although the problems of \cite{Tan08,Pirandola11,Nair11} naturally include various degrees of loss, few general results including its effects are available.  In this Letter, we first set up a general framework for image sensing in the presence of loss that subsumes many of the above problems.  We then identify a class of input states that contains an optimal, i.e., cost-minimizing, state for any problem fitting the framework, and under any form of signal energy constraint.
\begin{figure}
  \begin{center}
   \begin{tabular}{c}
   \includegraphics[trim=20mm 226mm 110mm 20mm, clip=true,width=0.5\textwidth]{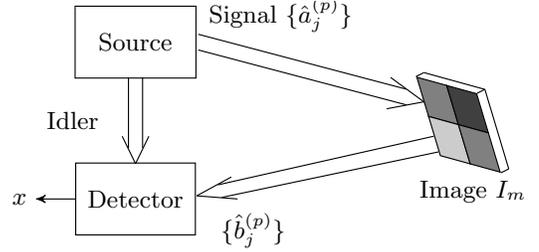}
   \end{tabular}
   \end{center}
\caption{\small Schematic of procedure for sensing of an unknown image from a set $\cl{I}$ with pixels described by $(\eta_m^{(p)},\theta_m^{(p)})$, $p = 1,\dotsc,P$, via \eqref{beamsplitter}. The source generates signal modes $\{\hat{a}_j^{(p)}\}$ for
probing the image and idler modes which are retained losslessly.
An optimal measurement for $x \in \cl{X}$ is made on the $J$ return modes
$\{\hat{b}_j^{(p)}\}$ and $J'$ idler modes jointly.}
\end{figure}

\emph{General Image Sensing Framework}: Suppose an image is drawn, unknown to the receiver, from a set $\cl{I}=\{I_1, \ldots, I_M\}$ of $M$ images according to the probability distribution $\{\pi_1, \ldots, \pi_M\}$. We model each image as a pixelated (transmissive or reflective) optical mask with uniform transmissivity/reflectivity and phase shift within each pixel. For $P$ the number of pixels in each image, the $p$-th pixel ($p \in \{1, \ldots, P\}$) of image $I_m$ is modeled as a beam splitter effecting the mode transformation
\begin{align} \label{beamsplitter}
\begin{pmatrix}
\hat{b}_j^{(p)} \\
\hat{f}_j^{(p)}
\end{pmatrix} =
\begin{pmatrix}
\sqrt{\eta_m^{(p)}}  e^{i\theta_m^{(p)}} & \sqrt{1 - \eta_m^{(p)}} \\
\sqrt{1 - \eta_m^{(p)}} e^{i\theta_m^{(p)}} & -\sqrt{\eta_m^{(p)}}
\end{pmatrix}
\begin{pmatrix}
\hat{a}_j^{(p)} \\
\hat{e}_j^{(p)}
\end{pmatrix}.
\end{align}
Here $\eta_m^{(p)}$ is the transmittance (or reflectance if reflective probing is used) of the $p$-th pixel in $I_m$, and $\theta_m^{(p)}$ is the phase shift imparted to the input (or ``signal'') field modes probing the $p$-th pixel of $I_m$ alone. 

For probing the unknown image, we consider quantum states of $J = \sum_{p=1}^P J^{(p)}$ signal modes that may be entangled to ancilla (or ``idler'') modes that are not sent out to interrogate the image but held losslessly (see Fig.~1). Here, $J^{(p)}$ is the number of modes interrogating pixel $p$, which, for example, could be successive time modes. Eq.~\eqref{beamsplitter} includes the annihilation operator $\hat{a}_j^{(p)}$ of the $j$-th signal field mode probing pixel $p$ and the annihilation operator $\hat{e}_j^{(p)}$ of the $j$-th input environment mode at $p$. The input environment modes are taken to be in the vacuum state -- the assumption of no thermal noise in the environment is realistic at optical frequencies. We further assume that the output mode corresponding to $\hat{b}_j^{(p)}$ in \eqref{beamsplitter}, but not that corresponding to $\hat{f}_j^{(p)}$, is available for making quantum measurements. This is almost always the case in practice as the environment input and output modes are not easily accessible to the user in a standoff imaging scenario, or if the light source and receiver are at different spatial locations. Additional loss during state propagation may be included as a multiplicative factor in the $\{\eta_m^{(p)}\}$.

We first consider pure input states -- we will return to the mixed input state case later.
An arbitrary pure quantum state $\ket{\psi}_{IS}$ of the signal and idler modes may be written in the form
\begin{align} \label{inputstate}
\ket{\psi}_{IS}= \sum_{\mbf{n}} c_{\mbf{n}} \keti{\phi_{\mbf{n}}} \kets{\mbf{n}}.
\end{align}
Here $\mbf{n} = (n^{(1)}_1, \ldots, n^{(1)}_{J^{(1)}}, \ldots, n^{(P)}_1, \ldots, n^{(P)}_{J^{(P)}})$ is a $J$-dimensional vector whose component $n_{j}^{(p)}$ indexes the photon number in the $j$-th mode interrogating the $p$-th pixel, and $\kets{\mbf{n}}$ are Fock states of the signal modes. We do not restrict the number $J'$ of the idler modes, nor the form of the idler states $\keti{\phi_{\mbf{n}}}$, only requiring that they be normalized. Allowing an arbitrary input state \eqref{inputstate} corresponds to the most general \emph{ancilla-assisted parallel strategy} (see \cite{GLM11} for discussion on parallel strategies). Little is known about non-parallel strategies (but see \cite{optimalstrategies1,*optimalstrategies2}) that may include adaptive selection of inputs which is known to assist some channel discrimination problems \cite{Harrow10}.
Irrespective of the form of the $\{\keti{\phi_{\mbf{n}}}\}$, the probability mass function (pmf) of the photon number in the signal modes is $p_{\mbf{n}} = |c_{\mbf{n}}|^2$, which determines quantities of interest such as the mean total signal energy:-
\begin{align} \nonumber
\label{signalenergy}
  \left\langle  \sum_{p=1}^{P} \sum_{j=1}^{J^{(p)}} \hat{N}^{(p)}_j \right\rangle
  = \sum_{\mbf{n}} n p_{\mbf{n}},
\end{align}
where
\begin{align} \nonumber
  n= \sum_{p=1}^{P} \sum_{j=1}^{J^{(p)}} n_{j}^{(p)} \equiv \sum_{p=1}^{P} n^{(p)}.
\end{align} In practice, the mean total signal energy may be upper bounded by a given number $N_S$.

Once an input state $\ketsi{\psi}$ is chosen and the signal modes are sent to probe the image, the return+idler states constitute an ensemble  $\cl{E}=\{(\pi_m,\rho_m)\}_{m = 1}^M,$ where $\rho_m= \mathrm{id}_I \otimes \cl{K}_m (\ket{\psi}_{IS}\bra{\psi})$ is the density operator on the return+idler Hilbert space at the output of the quantum channel $\mathrm{id}_I \otimes \cl{K}_m$ resulting from the interaction of the signal modes with $I_m$ via \eqref{beamsplitter} and the identity map on the idler modes. Depending on the imaging task, we attempt to extract a parameter lying in an observation space $\cl{X}$ by making a quantum measurement that is represented by a POVM \cite{helstrom76} with outcomes $x \in \cl{X}$ and corresponding operators $\{E_{x}\}_{x \in \cl{X}}$.  The task also specifies a \emph{cost function} $C(m,x)$, and we are interested in the \emph{minimum average cost} $\overline{C}$
 \begin{align}
 \ovl{C}[\cl{E}] = \begin{array}{c} \min  \\ \{E_{x}\} \end{array} \sum_{x \in \cl{X}}\sum_{m=1}^{M} \pi_m \tr (\rho_m E_{x}) C(m,x),
\end{align}
where the minimization is over all POVMs $\{E_x\}_{x \in \cl{X}}$. Therefore, adaptive \emph{measurements} are included in our model. Note that the input state and image parameters enter into the cost via the ensemble $\cl{E}$ while the imaging task determines $\cl{X}$ and the cost function. Thus, choosing $\cl{X}= \{1, \ldots, M\}$ and $C(m,x)=1-\delta_{m,x}$ makes $\ovl{C}[\cl{E}]$ equal the minimum probability of error (MPE) in discriminating the $M$ images. For the same cost function, $M=2$, $P=1$, and $\theta_m^{(1)} \equiv 0$ corresponds to the quantum reading and target detection problems of \cite{Tan08,Pirandola11,Nair11}. For $P=1$ and $\eta_m^{(1)} \equiv \eta$, choosing $\cl{X} = [0, 2\pi)$ and $C(m,x)= [x-\theta_m^{(1)}]^2$ corresponds to minimum-mean-square-error (MMSE) phase discrimination in the presence of loss. As $M \rightarrow \infty$, we approach MMSE phase estimation in loss. The usual interferometric setup for phase estimation \cite{DD1,GLM11} is recovered using $P=2$, $m= (\delta,\phi) \in [0,2\pi)^2$, $\theta_m^{(1)}=\phi + \delta$, $\theta_m^{(2)}= \phi$, and the cost function $C(m,x) = [x-\delta]^2$. Here $\delta$ is the relative phase shift of interest and $\phi$ is the undesired common phase shift in both arms of the interferometer (see Fig.~2).
\begin{figure}
  \begin{center}
   \begin{tabular}{c}
   \includegraphics[trim=15mm 243mm 110mm 15mm, clip=true,width=0.5\textwidth]{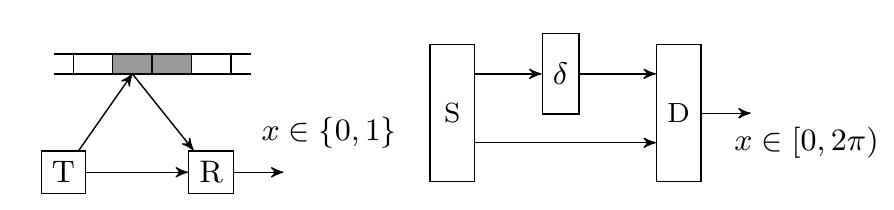}
   \end{tabular}
   \end{center}
   \caption{Image sensing problems in linear loss. Left: Quantum reading -- Digital reader composed of a transmitter $T$ and receiver $R$.  Right: Interferometer for discrimination/estimation of the relative phase shift $\delta$ with two-mode signal-only source $S$ and detector $D$.}
\end{figure}

In the state \eqref{inputstate}, if $\keti{\phi_{\mbf{n}}} \equiv \keti{\phi}$ for all $\mbf{n}$ and some idler state $\keti{\phi}$, the state factorizes so that the idler is effectively absent. The other extreme is the case of $\braket {\phi_{\mbf{n}}}{\phi_{\mbf{n'}}}_{I} = \delta_{{\mbf{n}},{\mbf{n'}}}$ so that the density operator on the signal modes is diagonal in the multimode Fock basis. Such states are called Number-Diagonal Signal (NDS) States in \cite{Nair11}, well-known examples being the two-mode squeezed vacuum of \cite{Tan08,Pirandola11} and the NOON state \cite{Dowling08}. In \cite{Nair11},  the error probability and other quantities of interest  were computed for NDS inputs in the $M=2$ case. Our main result is that the NDS states are optimally `matched' to the general imaging problem described above.
\begin{thm} [NDS State Lower Bound]
Let $\mathcal{I}$ be a set of $M$ $P$-pixel images described via transformations of the
form \eqref{beamsplitter} with prior probabilities $\{\pi_m\}_{m=1}^M$.
For any imaging task with cost function $C(m,x)$,
the minimum cost $\ovl{C}$ achieved by the input state $\ket{\psi}_{IS}= \sum_{\mbf{n}} c_{\mbf{n}} \keti{\phi_{\mbf{n}}} \kets{\mbf{n}}$ is lower
bounded by that achieved by a corresponding NDS state
$\ket{\Psi}_{IS}= \sum_{\mbf{n}} c_{\mbf{n}} \keti{\Phi_{\mbf{n}}} \kets{\mbf{n}}$, where $\{\keti{\Phi_{\mbf{n}}}\}$ is any orthonormal set, with the same signal photon pmf.
\end{thm}
\emph{General Results in Quantum Decision Theory}: The proof of Theorem 1 requires two simple but general results in quantum decision theory. To state them, we define the notion of \emph{mixture of ensembles}:- For each value of an
arbitrary index $l$ with associated probability $\lambda_l$, let
$\cl{E}_l = \{(\pi_m^{(l)},\rho_m^{(l)})\}_{m=1}^M$ be an $M$-ary ensemble
of states in a Hilbert space $\cl{H}$.   Then the $M$-tuple of pairs
\begin{equation} \label{mixture}
   \cl{E}= \sum_l \lambda_l \cl{E}_{l} := \Biggl\{\biggl( \sum_l \lambda_l \pi_m^{(l)}, \frac{\sum_l \lambda_l \pi_m^{(l)} \rho_m^{(l)}} {\sum_l \lambda_l \pi_m^{(l)}} \biggr)\Biggr\}_{m=1}^M
\end{equation}
is also an ensemble, called the \emph{mixture of the ensembles $\{\cl{E}_{l}\}$} -- the $\{\cl{E}_{l}\}$ are \emph{sub-ensembles of} $\cl{E}$. A mixture of ensembles can arise from a two-step procedure in which  $l$ is chosen with probability $\lambda_l$, following which the state $\rho_m^{(l)}$ is prepared with probability $\pi_m^{(l)}$. In our proof of Theorem 1, sub-ensembles arise as the conditional output states of a measurement with outcomes $\{l\}$ on a given ensemble. We have the following basic result: 
\begin{lemma} [Concavity of $\ovl{C}$ under mixing of ensembles]
Consider a sensing task with cost function $C(m,x)$. For $M$-ary ensembles $\{\cl{E}_{l}\}$ indexed by $l$, and probability distribution $\{\lambda_l\}$,
\begin{align} \label{concavityofCbar}
   \ovl{C}\biggl[\sum_l \lambda_l \cl{E}_{l}\biggr]
   \geq \sum_l \lambda_l \ovl{C}\left[\cl{E}_{l}\right].
\end{align}
\end{lemma}
\noindent The notion of \emph{orthogonal ensembles} provides a sufficient condition for equality in \eqref{concavityofCbar}. The \emph{support of an ensemble} $\cl{E}=\{(\pi_m,\rho_m)\}_{m = 1}^M$ is defined to be  $\supp \cl{E} := \sum_{m=1}^{M} \ran \rho_m$, where $\ran \rho_m$  is the range of $ \rho_m$. Then ensembles $\cl{E}$ and $\cl{F}$ are said to be \emph{orthogonal} if the support spaces $\supp \cl{E}$ and $\supp \cl{F}$ are orthogonal.
\begin{lemma}
If $\{\cl{E}_{l}\}$ are pairwise orthogonal ensembles on $\cl{H}$, we have
\begin{align} \label{equality}
    \ovl{C}\biggl[\sum_l \lambda_l \cl{E}_{l}\biggr]
  = \sum_l \lambda_l \ovl{C}\left[\cl{E}_{l}\right].
\end{align}
\end{lemma}
\noindent The proofs of Lemma 1 and 2 are given in 
the Appendix along with their physical interpretation. 

\emph{Proof of Theorem 1:} Suppose the general state \eqref{inputstate} is used as input.  To calculate the output states $\{\rho_m\}_{m=1}^M$, we may use the Schr\"{o}dinger-picture form of \eqref{beamsplitter} to get the purification
\begin{align}
\label{psim}
  \ket{\psi_m}
   &= \sum_{\mbf{l}}
      \Biggl( \sum_{\mbf{n} \geq \mbf{l}} c_{\mbf{n}} A_m^{(\mbf{n}:\mbf{l})} \keti{\phi_{\mbf{n}}}   \kets{\mbf{n}-\mbf{l}}
      \Biggr)
      \kete{\mbf{l}} \\
   &\equiv  \sum_{\mbf{l}} \ket{\psi_m^{(\mbf{l})}}_{IS} \kete{\mbf{l}},
\end{align}
where $\kete{\mbf{l}}$ is a Fock state of the environment modes and 
\begin{align}
   &A_m^{(\mbf{n}:\mbf{l})} = \nonumber \\
  & \prod_{p=1}^P   \left[ e^{i\theta_m^{(p)}n^{(p)}}
\prod_{j=1}^{J^{(p)}} \sqrt{ \binom{n_j^{(p)}}{l_j^{(p)}} {\eta_m^{(p)}}^{n_j^{(p)} - l_j^{(p)}} (1-\eta_m^{(p)})^{l_j^{(p)}}}\right].\nonumber
\end{align}
The output state $\rho_m$ is then given by (note that $\ket{\psi_m^{(\mbf{l})}}_{IS}$ are non-normalized states):
\begin{align}\label{rhom}
\rho_m = \sum_{\mbf{l} } \ket{\psi_m^{(\mbf{l})}}_{IS} \bra{\psi_m^{(\mbf{l})}}.
\end{align}
In (\ref{psim}-\ref{rhom}), $\mbf{l}$ is the (random and unknown) pattern of the number of photons leaked from the signal modes into the environment modes during interrogation of the image. The probability that the leaked photon pattern is $\mbf{l}$ is
\begin{align} \label{lambdal}
\lambda_{\mbf{l}} = \sum_{m=1}^{M} \pi_m \braket{\psi_m^{(\mbf{l})}}{\psi_m^{(\mbf{l})}}_{IS}
\end{align}
so that the conditional probability of hypothesis $m$ given $\mbf{l}$ is 
\begin{align}
\label{piml}
  \pi_m^{(\mbf{l})}
  = \frac{\pi_m \braket{\psi_m^{(\mbf{l})}}{\psi_m^{(\mbf{l})}}_{IS}}
         {\lambda_{\mbf{l}}}.
\end{align}
Thus, the ensemble \begin{align} \label{imagemixture} \cl{E} = \{(\pi_m, \rho_m)\}_{m=1}^M = \sum_{\mbf{l}} \lambda_{\mbf{l}} \cl{E}_{\mbf{l}} \end{align} for the sub-ensembles $\{\cl{E}_{\mbf{l}}\}$ given by
\begin{align}
\cl{E}_{\mbf{l}} = \left\{\left(\pi_m^{(\mbf{l})}, \frac{\ketsi{\psi_m^{(\mbf{l})}}\brasi{\psi_m^{(\mbf{l})}} } {\langle\psi_m^{(\mbf{l})}\ketsi{\psi_m^{(\mbf{l})}}}\right) \right\}_{m=1}^M.\end{align}
According to Lemma 1, the mixture \eqref{imagemixture} satisfies
\begin{align} \label{imageensembleconcavity}
\ovl{C}\left[\cl{E}\right] \geq \sum_{\mbf{l}} \lambda_{\mbf{l}}\ovl{C}\left[\cl{E}_{\mbf{l}}\right].
\end{align}
Consider the RHS of \eqref{imageensembleconcavity}. For each $\mbf{l}$, $\cl{E}_{\mbf{l}}$ is a pure-state ensemble, so $\ovl{C}[\cl{E}_{\mbf{l}}]$ is a function of just the pairwise inner products  $\braket{\psi^{(\mbf{l})}_m}{\psi^{(\mbf{l})}_{m'}}_{IS} \equiv G_{m,m'}^{(\mbf{l})}$, the $M\times M$ \emph{Gram matrix} 
\footnote{Two pure-state ensembles with the same prior probabilities and the same Gram matrix have the same $\ovl{C}$ because there exists a unitary $\hat{U}$ connecting the two ensembles (and also the POVMs on the two ensembles). This is seen by performing Gram-Schmidt orthogonalization on the two state sets separately -- $\hat{U}$ is chosen so as to map one Gram-Schmidt basis to the other.}:-
\begin{align} \label{graml}
G_{m,m'}^{(\mbf{l})} = \sum_{\mbf{n} \geq \mbf{l}} p_{\mbf{n}}  A_m^{(\mbf{n}:\mbf{l})*} A_{m'}^{(\mbf{n}:\mbf{l})}.
\end{align}
The crucial point is that, owing to the form \eqref{psim} of the beam splitter transformation,  $G_{m,m'}^{(\mbf{l})}$ is \emph{independent of the choice of the} $\{\keti{\phi_{\mbf{n}}}\}$. From (\ref{lambdal}-\ref{piml}), so are $\lambda_{\mbf{l}}$ and $\pi_m^{(\mbf{l})}$. Thus, the hypothetical measurement scenario in which one has knowledge of  $\mbf{l}$ (or alternatively, one is allowed to make a photon number measurement on all the output environment modes) and whose $\overline{C}$ is given by the RHS of \eqref{imageensembleconcavity}, has the \emph{same $\overline{C}$ for any choice of the} $\{\keti{\phi_{\mbf{n}}}\}$.

Finally, we consider the NDS input state $\ket{\Psi}_{IS}= \sum_{\mbf{n}} c_{{\mbf{n}}} \keti{\Phi_{\mbf{n}}} \kets{\mbf{n}}$ corresponding to \eqref{inputstate} satisfying
\begin{align} \label{NDS}
\braket {\Phi_{\mbf{n}}}{\Phi_{\mbf{n'}}}_{I} = \delta_{\mbf{n},\mbf{n'}}.
\end{align}
It is readily verified using (\ref{psim}) and \eqref{NDS} that
\begin{align} \label{NDSorthogonalensemble}
\braket{\psi_m^{(\mbf{l})}}{\psi_{m'}^{(\mbf{l'})}}_{IS} = \delta_{\mbf{l}, \mbf{l'}} \hspace{1mm}
\braket{\psi_m^{(\mbf{l})}}{\psi_{m'}^{(\mbf{l})}}_{IS}
\end{align}
so that the $\{\cl{E}_{\mbf{l}}\}$ are pairwise orthogonal ensembles. Therefore, by Lemma 2, the NDS input $\ket{\Psi}_{IS}$ \emph{attains} $\sum_{\mbf{l}} \lambda_{\mbf{l}}\ovl{C}\left[\cl{E}_{\mbf{l}}\right]$, and does so with the same signal photon pmf as $\ket{\psi}_{IS}$.

\emph{Discussion and Implications}: The $\ovl{C}$ of the NDS state $\ketsi{\Psi}$ of Theorem 1 is a function of only the $J$-mode photon pmf $\{p_{\mbf{n}}\}$. Thus, for a given $J$, the search for an optimal input state for a given imaging task may be confined to the set of $\{p_{\mbf{n}}\}$ satisfying given constraints, e.g., an average/peak signal energy constraint or a mode-by-mode signal energy constraint. As illustrated below, the problem of finding the optimal quantum state reduces to the classical problem of finding an optimal probability distribution.  To see that mixed input states $\rho_{IS}$ do not help, we first purify $\rho_{IS}$ using added idler modes. As \eqref{inputstate} contains no restriction on the idlers,  the NDS state corresponding to the purification -- which has the same $\{p_{\mbf{n}}\}$ as $\rho_{IS}$ -- has $\ovl{C}$ not larger than that of the purification (which in turn beats $\rho_{IS}$). Note also that the performance achieved by an arbitrary state of signal energy $N_S$ can be achieved by an NDS state of total (signal+idler) energy not larger than $2N_S$ by choosing $\keti{\phi_{\mbf{n}}} = \keti{\mbf{n}}$, the Fock state of the idler modes.

Theorem 1 strongly suggests that ancilla-assisted parallel strategies for image sensing are superior to signal-only parallel strategies. This is known for discrimination between some pairs of channels \cite{J'>0helpsbinarychanneldiscrimination} and is also true for our phase discrimination example below. We conjecture that they are strictly better whenever nonzero loss is present ($\eta_m^{(p)} <1$) because for a signal-only input state, the $\{\cl{E}_{\mbf{l}}\}$ are not orthogonal and are unlikely to achieve the lower bound of \eqref{imageensembleconcavity}. Such ancilla-assisted schemes appear to be unexplored for some problems of interest -- e.g., studies of the optimal state for phase estimation in loss have hitherto been confined to two-mode signal-only states \cite{DD1}. At the same time, Theorem 1 implies that the best possible performance can be obtained without ancillary modes if the value of $\mbf{l}$ is known. This result should place interesting limitations on the quantum advantage obtainable in any sensing problem.

\emph{Binary Phase Discrimination}: As an application of Theorem 1, we obtain the single-pass ($J=1$) state that discriminates between a $0$ and $\pi$ phase shift with minimum error probability among states with a peak signal photon constraint of $N_\textrm{peak} = 2$ in the presence of loss. In the terminology of our framework, we have $M=2$, $P=1$, $\eta_1^{(1)} = \eta_2^{(1)} = \eta <1$, and $\theta_1^{(1)} =0, \; \theta_2^{(1)} = \pi$. We also assume $\pi_1=\pi_2 =1/2$. An arbitrary state  $\ketsi{\psi}$ satisfying these constraints is
\begin{align} \label{N=2state}
\ketsi{\psi} = \sqrt{p_0}\keti{\phi_0}\kets{0}+\sqrt{p_1}\keti{\phi_1} \kets{1}+ \sqrt{p_2}\keti{\phi_2}\kets{ 2} ,
\end{align}
where phase factors have been absorbed into the normalized kets $\{\keti{\phi_n}\}$. In terms of the density operators $\rho_1$ and $\rho_2$ defined earlier, the minimum error probability is given by the Helstrom formula \cite{helstrom76}
\begin{align}\label{helstrom}
\overline{P}_e= \frac{1}{2} -\frac{1}{4} ||\rho_1- \rho_2||_1,
\end{align} 
where $||\cdot||_1$ is the trace norm. According to Theorem 1, we may confine our search for optimal states to the NDS class for which $\{\keti{\phi_n}\}$ are orthonormal. For such states, the minimum error probability is given in closed form by Eq.~(39) of \cite{Nair11}:
\begin{align} \nonumber
\overline{P}_e^{\tsf{NDS}} = {1/2} -\sqrt{p_1} \left[(p_0\eta+p_2 \eta^3)^{1/2} + (2p_2\eta(1-\eta)^2)^{1/2}\right].
\end{align}
Since $p_0+p_1+p_2=1,$ we may consider $p_0$ and $p_1$ as independent variables taking values in the triangle $\cl{T}$ whose vertices have the $(p_0,p_1)$ values  $(0,0), (1,0),$ and $(0,1)$. It is easy to show that $\overline{P}_e^{\tsf{NDS}}$ is identically $1/2$ on the $p_0-$axis and that it has local minima on the $p_1$-axis at $p_1=p_2=1/2$ and on the remaining boundary of the triangle at $p_0=p_1=1/2$. There also exists a local extremum of $\overline{P}_e^{\tsf{NDS}}$ in the interior of $\cl{T}$ at the point
\begin{align} \nonumber 
(p_0^{*},p_1^*, p_2^*) &= \left(\frac{1+2\eta-\eta^2}{2(1+\eta)(3-\eta)}, \frac{1}{2}, \frac{1} {(1+\eta)(3-\eta)}\right). 
\end{align}
Fig.~3 (left panel) shows $\overline{P}_e^{\;\tsf{NDS}}$  plotted over $\cl{T}$ for $\eta=0.6$. The interior extremum point $(p_0^*,p_1^*,p_2^*)$ achieves the minimum error probability. For comparison, we consider also a signal-only input state of the form 
\begin{align} \label{N=2signalonly}
\kets{\psi} = \sqrt{p_0}\kets{0}+\sqrt{p_1} \kets{1}+ \sqrt{p_2}\kets{ 2}.
\end{align}
For each choice of $(p_0,p_1)$ in $\cl{T}$, the error probability $\overline{P}_e^{\tsf{signal-only}}$ is computed numerically using \eqref{helstrom}, and in Fig.~3 (right panel), the difference $\overline{P}_e^{\tsf{signal-only}}-\overline{P}_e^{\tsf{NDS}}$ is plotted on $\cl{T}$. The difference is everywhere non-negative, being zero on the two boundaries of $\cl{T}$ other than the $p_1-$axis.

\begin{figure}
  \begin{center}
   \begin{tabular}{c}
   \includegraphics[trim=35mm 137mm 130mm 65mm, clip=true,width=0.5\textwidth]{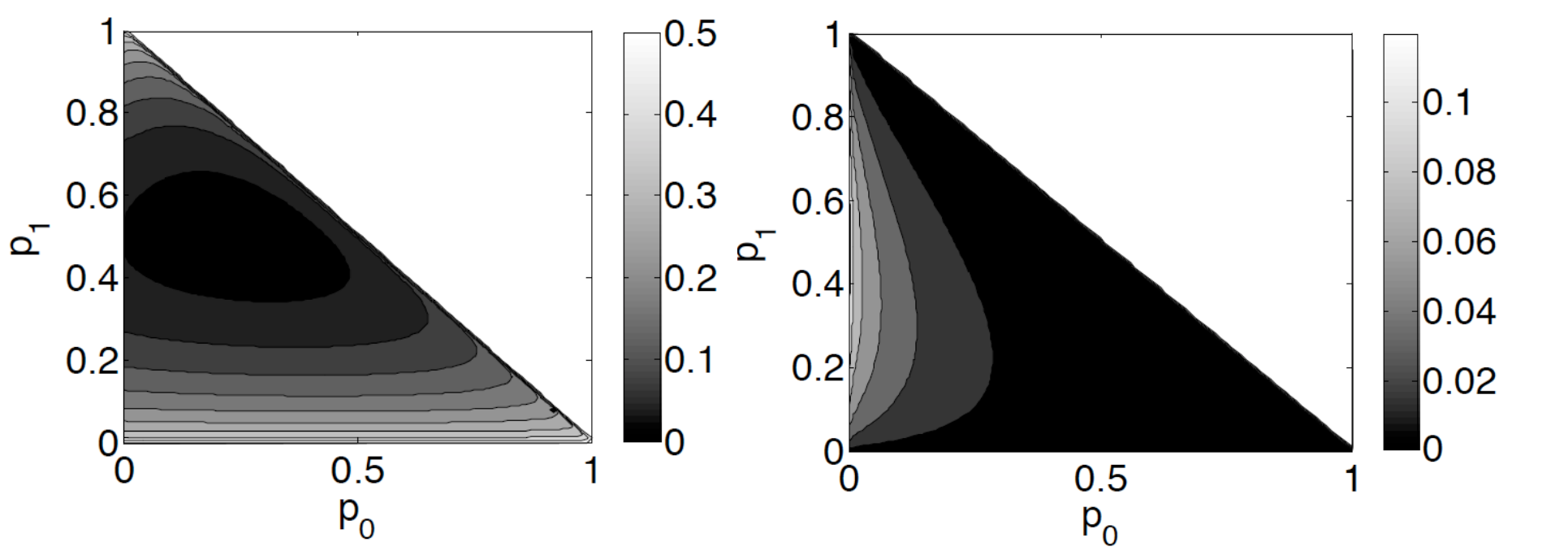}
   \end{tabular}
   \end{center}
   \caption{\label{fig:3} Left: The error probability $\overline{P}_e^{\tsf{NDS}}$ of the NDS state of the form of Eq.~\eqref{N=2state} for $\eta=0.6$. Right:  The difference $\overline{P}_e^{\tsf{signal-only}}-\overline{P}_e^{\tsf{NDS}}$ between the error probabilities of corresponding signal-only and NDS states as a function of $(p_0,p_1)$.}
 \end{figure}

\emph{Lossless Image Sensing}: In the lossless case, $\mbf{l}=\mbf{0}$ with probability one, so that the performance of the hypothetical measurement described in Theorem 1 is attainable with \emph{any} choice of idler states. That performance is determined by the Gram matrix elements from \eqref{graml}:-
\begin{align}
  G_{m,m'}^{(\mbf{0})}
  &= \sum_{\mbf{n} }
      \Biggl(p_{\mbf{n}}
      \prod_{p=1}^P e^{i\left(\theta_{m'}^{(p)}-\theta_m^{(p)}\right)n^{(p)}}
      \Biggr)  \nonumber \\
  &\equiv \sum_{\mbf{\nu}} p_{\mbf{\nu}} \hspace{1mm} e^{i\sum_{p=1}^P \left(\theta_{m'}^{(p)}-\theta_m^{(p)}\right)n^{(p)}} \nonumber \end{align}
where $\nu=\left(n^{(1)}, \ldots, n^{(P)}\right) $ has pmf $p_{\nu}$. Choosing $\keti{\phi_{\mbf{n}}} \equiv \keti{\phi}$, the signal-only state
\begin{align}
\kets{\psi} = \sum_{\nu} \sqrt{p_{\nu}} \kets{n^{(1)}, \ldots, n^{(P)}} \nonumber
\end{align}
with $J^{(p)}=1, J'=0$ suffices to attain $\ovl{C}$. In the absence of loss, the $\{\cl{K}_m\}_{m=1}^M$ are unitary channels. This result shows that, among parallel strategies, ancillas do not improve sensing of $M$ unitary phase images under a signal energy constraint. This is unlike the case of minimum error probability discrimination of finite-dimensional unitaries in \cite{J'>0helpsM-aryunitarydiscrimination}, although ancillas are not required for discriminating two unitaries \cite{J'>0helpsM-aryunitarydiscrimination,CPR00}. The fact that single-pass imaging ($J^{(p)}=1$) suffices is also remarkable as there are examples of pairs of unitaries that are better (even perfectly) discriminated if multiple shots are allowed \cite{J'>0helpsM-aryunitarydiscrimination,J>1helpsbinaryunitarydiscrimination}.

We acknowledge useful discussions with Masoud Mohseni and Jeffrey H.~Shapiro. This work was supported by DARPA's Quantum Sensor Program under AFRL Contract No. FA8750-09-C-0194.
 

%

\section{Appendix: Proofs of Lemmas 1 and 2}

\noindent \textbf{Lemma 1} (Concavity of $\ovl{C}$ under mixing of ensembles). \textit{Consider an imaging task with cost function $C(m,x)$. For $M$-ary ensembles $\{\cl{E}_{l}\}$ indexed by $l$, and probability distribution $\{\lambda_l\}$,}
\begin{align} \label{concavityofCbar}
   \ovl{C}\biggl[\sum_l \lambda_l \cl{E}_{l}\biggr]
   \geq \sum_l \lambda_l \ovl{C}\left[\cl{E}_{l}\right].
\end{align}

\noindent \textit{Proof}. The LHS of \eqref{concavityofCbar} evaluates to
\begin{align}
    &\begin{array}{c} \min  \\ \{E_x\} \end{array} \sum_l \lambda_l \sum_{x \in \cl{X}}\sum_{m=1}^{M} \pi_m^{(l)} \tr \left(\rho_m^{(l)} E_x\right) C(m,x) \label{singlechoiceofPOVM}\\
  &\geq  \sum_l \lambda_l \left( \begin{array}{c} \min  \\ \{E_x^{(l)}\} \end{array}  \sum_{x \in \cl{X}}\sum_{m=1}^{M} \pi_m^{(l)} \tr \left(\rho_m^{(l)} E_x^{(l)}\right) C(m,x) \right) \label{multiplechoiceofPOVM} \\
   &= \sum_l \lambda_l \ovl{C}\left[\cl{E}_{l}\right]. \nonumber
\end{align}
The inequality \eqref{multiplechoiceofPOVM} follows because the inner sum is minimized separately for each value of $l$ in \eqref{multiplechoiceofPOVM} but not in \eqref{singlechoiceofPOVM}. $\Box$
\newline \newline
Physically, the RHS of \eqref{concavityofCbar} is the minimum cost when one knows the sub-ensemble (indexed by $l$) prior to making the measurement. Then \eqref{concavityofCbar} simply asserts that the cost is higher in the case where information on $l$ is not available.
\\ \\ \\ \\ \\ \\ 
\noindent \textbf{Lemma 2} \textit{If $\{\cl{E}_{l}\}$ are pairwise orthogonal ensembles on $\cl{H}$, we have}
\begin{align} \label{equality}
    \ovl{C}\biggl[\sum_l \lambda_l \cl{E}_{l}\biggr]
  = \sum_l \lambda_l \ovl{C}\left[\cl{E}_{l}\right].
\end{align}

\noindent \textit{Proof}. Let $\cl{H}_l = \supp \cl{E}_{l}$ for each $l$.  It is clear that it is suffices to consider POVMs on the subspace $ \cl{H}' = \bigoplus_l \cl{H}_l \subseteq \cl{H}$. For each value of $l$, let $\Pi_l$ denote the projection operator onto $\cl{H}_l$.  Since the $\{\cl{E}_{l}\}$ are pairwise orthogonal, we have $\Pi_{l} \Pi_{l'} = \delta_{l,l'} \Pi_{l}$. Let $\{E_x^{(l)}\}_{x \in \cl{X}}$ denote the POVM (on $\cl{H}'$) that realizes $\ovl{C}\left[\cl{E}_{l}\right]$. We define a POVM $\{E_x\}_{x \in \cl{X}}$ on $\cl{H}'$ with elements:-
\begin{align} \label{Ex}
E_x = \sum_l  E_x^{(l)} \Pi_l, \hspace{3mm}x \in \cl{X}.
\end{align}
The average cost realized by this POVM on the mixture $\cl{E} = \sum_l \lambda_l \cl{E}_{l}$ is then given by
\begin{align}
   & \sum_{x \in \cl{X}}\sum_{m=1}^{M}
     \tr \Biggl(\sum_{l,l'} \lambda_l \pi_m^{(l)} \rho_m^{(l)} E_x^{(l')} \Pi_{l'}\Biggr) C(m,x) \nonumber\\
   &=  \sum_{x \in \cl{X}}\sum_{m=1}^{M}
     \tr \Biggl(\sum_{l} \lambda_l  \pi_m^{(l)} \rho_m^{(l)} E_x^{(l)}\Biggr) C(m,x) \label{byorthogonality} \\
   &= \sum_{l} \lambda_l \ovl{C}\left[\cl{E}_{l}\right] \label{bydefinition}.
\end{align}
Here, \eqref{byorthogonality} follows from
\begin{align} \label{stateorthogonality}
\Pi_{l'} \rho_m^{(l)}  = \rho_m^{(l)}\cdot \delta_{l,l'},
\end{align}
which in turn follows from the orthogonality of the ensembles $\{\cl{E}_{l}\}$. Equality \eqref{bydefinition} follows from the definition of $\{E_x^{(l)}\}$. The fact that the POVM $\{E_x\}$ achieves the RHS of \eqref{concavityofCbar} establishes the claim \eqref{equality}. $\Box$

Physically, the optimum measurement on $\cl{E}$ may be regarded as a two-step measurement that learns the value of $l$ first and then makes the measurement achieving $\ovl{C}[\cl{E}_{l}]$. The orthogonality of the $\{\cl{E}_{l}\}$ ensures that the first measurement can be performed perfectly without adding noise to the second.

\end{document}